\documentclass[notitlepage,showkeys,11pt,noeprint,superscriptaddress]{revtex4-1}
\usepackage{graphicx}
\usepackage{subfigure}
\usepackage{amsmath}
\usepackage{amsfonts}
\usepackage{amssymb}

\usepackage[colorlinks=true, citecolor=cyan]{hyperref}

\begin{document}

\title{Entropic uncertainty relations and quantum
Fisher information of top quarks in a large hadron collider}
\author{Biao-Liang Ye}
\email{biaoliangye@gmail.com}
\affiliation{Quantum Information Research Center,
Shangrao Normal University, Shangrao 334001, China}
\affiliation{Jiangxi Province Key Laboratory of Applied Optical Technology,
Shangrao Normal University, Shangrao 334001, China}

\author{Li-Yuan Xue}
\affiliation{School of Physical Science and Intelligent Education,
Shangrao Normal University, Shangrao 334001, China}

\author{Zhi-Qiang Zhu}
\affiliation{School of Chemistry and Environmental Science,
Shangrao Normal University, Shangrao 334001, China}

\author{Dan-Dan Shi}
\affiliation{School of Mathematics and Computational Science,
Shangrao Normal University, Shangrao 334001, China}

\author{Shao-Ming Fei}
\email{feishm@cnu.edu.cn}
\affiliation{School of Mathematical Sciences, Capital Normal University, Beijing 100048, China}
\affiliation{Max-Planck-Institute for Mathematics in the Sciences, 04103 Leipzig, Germany}

\begin{abstract}
We employ the entropic uncertainty relations and the quantum Fisher information to explore the formation of quark $t\bar{t}$ pairs at a large hadron collider through the combination of $q\bar{q}$ pair and $gg$ pair initiated processes. A comprehensive analysis has been undertaken on the procedure of quark and gluon channel mixing in the production of top quark pairs $t\bar{t}$, encompassing the tightness of the entropic uncertainty inequalities and the maximum quantum Fisher information of the system.
\end{abstract}
\date{\today}
\maketitle

\section{Introduction}

The study on the standard model from the perspective of quantum information theory has resulted in a variety of rich results. The quark spin correlation of top quark pairs has been investigated at large hadron collider (LHC) \cite{Bernreuther1998, Bernreuther2004, Uwer2005, Baumgart2013, Bernreuther2015}. The quantum entanglement has been used to search for new physics in top quark pairs \cite{Cervera-Lierta2017, Afik2021, Aoude2022, Aguilar-Saavedra2023, Severi2023, Fabbrichesi2023, Aguilar-Saavedra2023a}. In addition, the Bell inequalities have been employed for examination \cite{Fabbrichesi2021, Afik2022, Severi2022, Barr2022, Barr2023} in $t\bar{t}$ events at LHC and Higgs boson \cite{Barr2022, Fabbrichesi2023}.

Establishing a connection between quantum information and high-energy physics is of great importance and has sparked numerous research endeavors. Recent reports have demonstrated that top quarks \cite{Afik2023} can effectively be characterized in terms of quantum entanglement \cite{Horodecki2009}, discord \cite{Ollivier2001} and steering \cite{Wiseman2007}. The top quark, the heaviest particle in the standard model, has a significant mass \cite{Afik2021} which contributes to a large decay width. The rapid decay of the top quark motivates researchers to use its unique characteristics to extract information about its spin from the decay products. Extensive investigations of the spin of top quarks, produced in top-antitop ($t\bar{t}$) pairs, have been conducted. The detection of spin correlations in top quarks is already a well-established technique at the Tevatron and LHC.

Among the various quantum features \cite{Modi2012, Adesso2016}, the quantum uncertainty principle is one of the essential issues in the quantum world. Originally formulated by Heisenberg, it clearly illustrates the distinction between the classical and quantum physics. Recently, the entropy-based uncertainty relation has been extended to encompass quantum entanglement \cite{Berta2010, Li2011} and utilized to detect quantum entanglement. On the other side, the quantum Fisher information is of crucial importance in quantum metrology \cite{Pezze2018}. It lies at the core of quantum metrology, providing a lower bound on the variance of an unbiased estimator. Larger quantum Fisher information implies smaller variance limit for the estimator, thereby enabling higher estimation precision. It has also been extensively used in the construction of entanglement witnesses \cite{Li2013}.

We study the top quark pairs from the perspective of entropic uncertainty relations and quantum Fisher information. For the specific case of $t\bar{t}$ pairs, we focus on the quantum state of the most elementary quantum chromodynamics (QCD) production processes involving interactions between light quark-antiquark ($q\bar{q}$) or gluon-gluon ($gg$). It is demonstrated that any realistic QCD process of $t\bar{t}$ production can be regarded as a statistical combination of $q\bar{q}$ and $gg$ at leading order. The manuscript is structured as follows. We introduce the top quark model and the basic notations and concepts of the entropic uncertainty relation and quantum Fisher information in Section II. We presented the main results in Section III. The study is concluded in Section IV.

\section{Preliminary}

\textbf{Top Quark-Antiquark Pair}
The primary focus of our research is the $t\bar{t}$ top quark pair. Top quarks are highly distinctive high-energy entities due to their measurable spin correlations. This study explores the overall structure of the quantum state of a top-antitop ($t\bar{t}$) quark pair generated through QCD within a high-energy collider. The examination of the quantum state of a $t\bar{t}$ pair generated via the fundamental QCD process necessitates an analysis of the entanglement and Bell inequalities \cite{Afik2022}. We study the entropic uncertainty and the quantum Fisher information related to the proton-antiproton collisions at  LHC and proton-proton collisions at the Tevatron.

Two spin-1/2 quarks, such as the $t\bar{t}$ pair, give rise to a two-qubit system. A $t\bar{t}$ pair is produced via high-energy proton-proton ($pp$) collisions in LHC. Protons consist of quarks (spin-1/2 fermions) and gluons (massless spin-1 bosons). Collectively, these constituents are referred to as partons. The interactions among these partons through quantum chromodynamics result in the production of a $t\bar{t}$ pair. For example, a $t\bar{t}$ pair can be generated through the interaction between a light quark and antiquark ($q\bar{q}$) or a pair of gluons (gg) as follows \cite{Afik2023,Afik2022,Afik2021},
$q+\bar{q}\rightarrow t + \bar{t}$ or $g+g\rightarrow t+\bar{t}$.
The production of a $t\bar{t}$ pair is described by the invariant mass $M_{t\bar{t}}$ and the direction $\hat{k}$ in the center-of-mass frame. Specifically, in this frame, the relativistic momenta of the top and antitop quarks are given by $k_t^\mu=(k_t^0,\mathbf{k})$ and $k_{\bar{t}}^\mu=(k_{\bar{t}}^0,-\mathbf{k})$, respectively, satisfying the invariant dispersion relation $k_t^2=m_t^2$ and $k_{\bar{t}}^2=k_t^2=m_t^2$.
The invariant mass is defined by these momenta,
\begin{equation}
M_{t\bar{t}}^2=s_{t\bar{t}}=(k_t+k_{\bar{t}})^2,
\end{equation}
where $s_{t\bar{t}}$ is the usual Mandelstam variable.
In the center-of-mass frame, this gives
$M_{t\bar{t}}^2=4(k_t^0)^2=4(m_t^2+\bf{k}^2)$.
Relating the momentum of the top quark to its velocity $\beta$, $|{\bf k}|=m_t\beta/\sqrt{1-\beta^2}$, we obtain
\begin{equation}
\beta=\sqrt{1-4m_t^2/M_{t\bar{t}}^2}.
\end{equation}
In the case of the threshold production ($\beta=0$), one has $M_{t\bar{t}}=2m_t\approx 346$ GeV, which represents the minimum energy required for the production of a $t\bar{t}$ pair.

The kinematics of the $t\bar{t}$ pair is given by the variables $(M_{t\bar{t}},\hat{t})$. For a fixed production process the spins of the $t\bar{t}$ pair are characterized by the following production spin density matrix $R(M_{t\bar{t}},\hat{k})$ \cite{Afik2021},
\begin{eqnarray}
R=\widetilde{A} I_4+\sum_i(\widetilde{B}_i^+
\sigma_i\otimes I_2+\widetilde{B}_i^-I_2\otimes\sigma_i)
+\sum_{i,j}\widetilde{C}_{ij}\sigma_i\otimes\sigma_j,
\end{eqnarray}
where the first/second spin subspace corresponds to the top/antitop, respectively. The production spin density matrix is parameterized by 16 parameters: $\widetilde{A}, \widetilde{B}_i^\pm, \widetilde{C}_{ij}$. The matrix $R$ is not properly normalized, with $tr(R)=4\widetilde{A}$, where $\widetilde{A}$ represents the differential cross-section for $t\bar{t}$ production at a fixed energy and top direction. The normalized form of the spin density matrix is given by
\begin{equation}
\rho=\frac14(I_4+\sum_i(B_i^+\sigma_i\otimes I_2+B_i^-I_2\otimes\sigma_i)+\sum_{i,j}C_{ij}\sigma_i\otimes\sigma_j),
\end{equation}
where the spin polarizations $B_i^\pm$ and spin correlations $C_{ij}$ of the $t\bar{t}$ pair can be computed according to $\rho=R/tr(R)=R/(4\widetilde{A})$, $B_i^\pm=\widetilde{B}_i^\pm/\widetilde{A}$ and $C_{ij}=\widetilde{C}_{ij}/\widetilde{A}$.

In theoretical computations the QCD perturbation theory is employed. At the leading order, two initial states can produce a $t\bar{t}$ pair: a $q\bar{q}$ pair or a $gg$ pair. Each of these initial states, represented by $I=q\bar{q}, gg$, results in a distinct quantum state for the $t\bar{t}$ pair when the energy and the top direction in the center-of-mass frame are fixed. The expression of $\rho_I$ is utilized to obtain
\begin{equation}
\rho(M_{t\bar{t},\hat{k}})=\sum_{I=q\bar{q},gg}
w_I(M_{t\bar{t}},\hat{k})\rho^I(M_{t\bar{t}},\hat{k}),
\end{equation}
where the probabilities $w_i$ are computed from the luminosities \cite{Afik2021}.
The production spin density matrix in the helicity basis is characterized by 5 parameters at the leading order. Specifically, for the $I=q\bar{q}$ process, the parameters are given by:
\begin{eqnarray}
\widetilde{A}^{q\bar{q}}&=&F_q(2-\beta^2\sin^2\Theta), \nonumber\cr
\widetilde{C}_{rr}^{q\bar{q}}&=&F_q(2-\beta^2)\sin^2\Theta, \nonumber\cr
\widetilde{C}_{nn}^{q\bar{q}}&=&-F_q\beta^2\sin^2\Theta, \nonumber\cr
\widetilde{C}_{kk}^{q\bar{q}}&=&F_q(2\cos^2\Theta+\beta^2\sin^2\Theta), \nonumber\cr
\widetilde{C}_{rk}^{q\bar{q}}&=&\widetilde{C}_{kr}^{q\bar{q}}
=F_q\sqrt{1-\beta^2}\sin2\Theta.
\end{eqnarray}
On the other hand, the parameters for the $I=gg$ process, the parameters are given by:
\begin{eqnarray}
\widetilde{A}^{gg}&=&F_g(\Theta)[1+2\beta^2\sin^2\Theta-\beta^4(1+\sin^4\Theta)], \nonumber\cr
\widetilde{C}_{rr}^{gg}&=&-F_g(\Theta)[1-\beta^2(2-\beta^2)(1+\sin^4\Theta)], \nonumber\cr
\widetilde{C}_{nn}^{gg}&=&-F_g(\Theta)[1-2\beta^2+\beta^4(1+\sin^4\Theta)], \nonumber\cr
\widetilde{C}_{kk}^{gg}&=&-F_g(\Theta)[1-\beta^2\sin^22\Theta/2-\beta^4(1+\sin^4\Theta)], \nonumber\cr
\widetilde{C}_{rk}^{gg}&=&F_g(\Theta)\sqrt{1-\beta^2}\beta^2\sin2\Theta\sin^2\Theta,
\end{eqnarray}
where the normalization factors are given by $F_q=1/18$ and $F_g(\Theta)=\frac{7+9\beta^2\cos^2\Theta}{192(1-\beta^2\cos^2\Theta)^2}$.

\textbf{Entropic Uncertainty Relations}
The entropic uncertainty relations have been experimentally demonstrated \cite{Li2011, Prevedel2011}. Maassen and Uffink presented the following entropic uncertainty relation for two observables $R$ and $S$ with eigenvectors $|a_i\rangle$ and $|b_j\rangle$, respectively,
\begin{eqnarray}
H(R)+H(S)\ge\log_2\frac1c,
\end{eqnarray}
where $H$ is the Shannon entropy and $c=\max_{i,j}|\langle a_i|b_j\rangle|^2$.

Berta et al. presented an entropic uncertainty relation satisfied by the measurement on system $A$ which is correlated with a system (quantum memory) $B$,
\begin{eqnarray}\label{eub}
H(R|B)+H(S|B)\ge\log_2\frac1c+H(A|B),
\end{eqnarray}
where $H(R|B)$ and $H(S|B)$ are the conditional von Neumann entropies, which quantify the uncertainty of the measurement outcomes of $R$ and $S$ given the information contained in $B$. $H(A|B)$ denotes the conditional von Neumann entropy between $A$ and $B$. As $-H(A|B)$ provides a lower bound on the one-way distillable entanglement, the lower bound of (\ref{eub}) is basically dependent on the entanglement between $A$ and $B$.

The von Neumann entropy of a density matrix $\rho_{AB}$ is defined by $H(AB):=-{\rm Tr}(\rho_{AB}\log_2\rho_{AB})=-\sum_i\lambda_i\log_2\lambda_i$, where $\lambda_i$ are the non-zero eigenvalues of $\rho_{AB}$. The conditional entropy $H(A|B)$ is given by $H(A|B)=H(AB)-H(B)$, where $H(B)$ is the von Neumann entropy of the reduced density operator $\rho_B=Tr_A\rho_{AB}$. Set $E_{L}=H(R|B)+H(S|B)$ and $E_{R}=\log_2\frac1c+H(A|B)$.
The tightness of the entropic uncertainty relation is given by $\Delta_E=E_L-E_R$.
The above entropic uncertainty relations have been experimentally demonstrated \cite{Li2011, Prevedel2011}. Similar to
concurrence \cite{Afik2021}, for Pauli matrices observables, quantum entanglement is witnessed if $E_R<1$.

\textbf{Quantum Fisher Information} The quantum Fisher information $F(\rho, A)$ of a quantum state $\rho$ with respect to an observable $A$ is defined by
\begin{eqnarray}
F(\rho, A)=\frac{1}{4}\mathrm{Tr}(\rho L^2),
\end{eqnarray}
where $L$ is the symmetric logarithmic derivative determined by
\begin{eqnarray}
i[\rho, A]=\frac{1}{2}(L\rho+\rho L),
\end{eqnarray}
with $[\cdot,\cdot]$ denoting the anti-commutator. For pure states, the quantum Fisher information reduces to the variance $V(\rho,A)$ of $\rho$ with respect to observable $A$,
\begin{eqnarray}
F(\rho, A)=V(\rho, A)= \mathrm{Tr}(\rho A^2)-(\mathrm{Tr}(\rho A))^2.
\end{eqnarray}

In general, when the state $\rho$ is mixed with spectral decomposition,
\begin{eqnarray}
\rho=\sum_{k=1}^d \lambda_k|k\rangle\langle k|,
\end{eqnarray}
where $\{|k\rangle\}$ are the eigenvectors of $\rho$, $d$ is the dimension of Hilbert space, the quantum Fisher information can be evaluated by using the following equation \cite{Braunstein1994, Li2013, Ye2018},
\begin{eqnarray}
F(\rho, A)=\sum_{k,l}\frac{(\lambda_k-\lambda_l)^2}
{2(\lambda_k+\lambda_l)}|\langle k|A|l\rangle|^2.
\end{eqnarray}

A set of observables $\{A_\mu\}$ is said to be a complete collection of orthonormal observables if ${\rm Tr} A_\mu A_\nu=\delta_{\mu\nu}$ and $\{A_\mu\}$ constitutes a basis of the associated space. For two-qubit systems, we consider the following complete collections of orthonormal observables $\{A_\mu\}$ and $\{B_\mu\}$ associated with systems $A$ and $B$, respectively,
\begin{eqnarray}
\{A_\mu\}=\{B_\mu\}=\{\frac{1}{\sqrt{2}},
 \frac{\sigma_1}{\sqrt{2}}, \frac{\sigma_2}{\sqrt{2}},
 \frac{\sigma_3}{\sqrt{2}}\},
\end{eqnarray}
where $\sigma_i$, $i=1,2,3$, are the standard Pauli matrices. The quantum Fisher information is given by
\begin{eqnarray}
F=\sum_\mu F(\rho^{ab}, A_\mu\otimes I^b+I^a\otimes B_\mu).
\end{eqnarray}
A state is separable if $F\leq2$. It has been shown that any state violating this inequality must be entangled \cite{Li2013}. Additionally, in this paper we consider $F$ to be the maximal value with respect to all unitary state preparations $U\rho^{ab}U^\dagger$, where $U\in \mathcal{U}(d)$ \cite{Fiderer2019}. Similar to the entanglement measure concurrence for top quarks at the LHC \cite{Afik2021}, quantum Fisher information can also serve as a witness of entanglement.

\section{Results and analysis}

We consider $t\bar{t}$ pair formed through the combination of $q\bar{q}$ pair and $gg$ pair initiated processes. We demonstrate the numerical results for the $t\bar{t}$ pair based on the entropic uncertainty relation and quantum Fisher information. The range of $\Theta$ is taken to be in $[0,\pi/2]$ and the $M_{t\bar{t}}$ varies between $346$ and $1000$ GeV.

\begin{figure}[htbp!]
\centering
\subfigure[]{
\includegraphics[width=.30\textwidth]{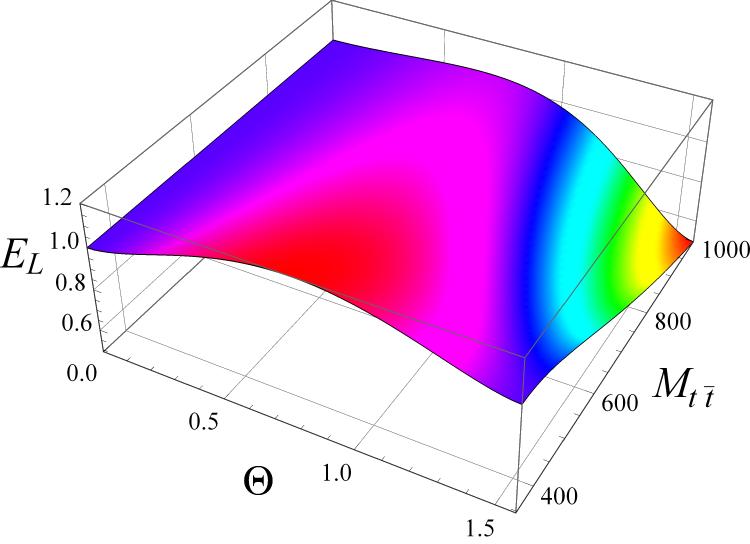}
}
\subfigure[]{
\includegraphics[width=.30\textwidth]{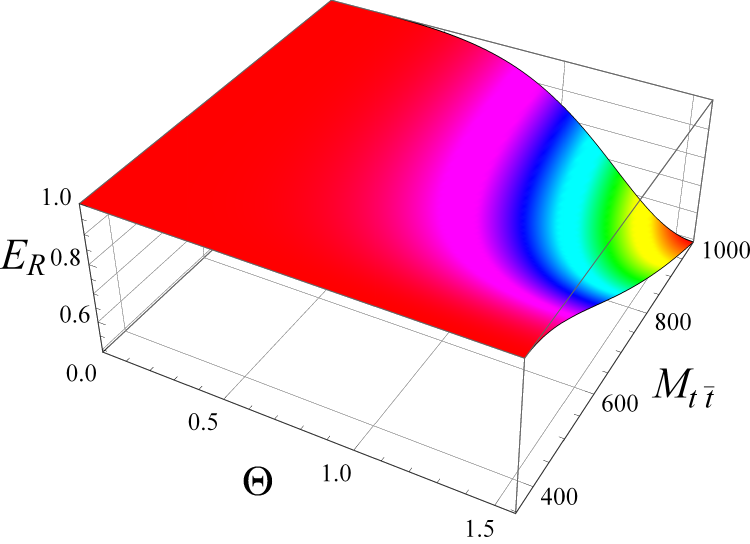}
}
\subfigure[]{
\includegraphics[width=.30\textwidth]{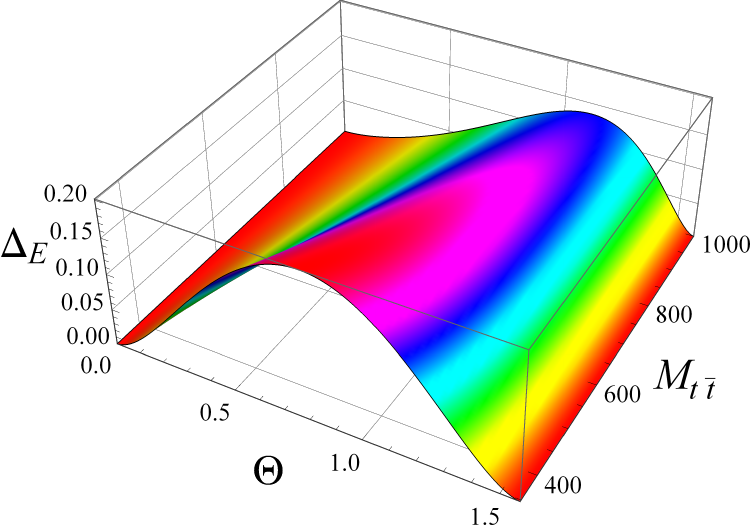}
}
\caption{The entropic uncertainty relation for the quarks $q\bar{q}$ production of a $t\bar{t}$ pair, $q\bar{q} \rightarrow t\bar{t}$. (a) $E_L$ with respect to $\Theta$ and $M_{t\bar{t}}$. (b), $E_R$ with respect to $\Theta$ and $M_{t\bar{t}}$. (c), $\Delta_E$ with respect to $\Theta$ and $M_{t\bar{t}}$. The unit of $M_{t\bar{t}}$ is GeV.
}
\label{f1}
\end{figure}

One observes the left-hand side of the entropic uncertainty relation, $E_{L}=H(R|B)+H(S|B)$ for $q\bar{q}$ in Fig. \ref{f1}. The observables are the Pauli matrices with $R=\sigma_x$ and $S=\sigma_z$. In Fig. 1(a), $E_L$ varies with the production angle $\Theta$ and the mass $M_{t\bar{t}}$. $E_L$ remains constant with $M_{t\bar{t}}$ when $\Theta=0$. As $\Theta$ increases, $E_L$ gradually decreases with $M_{t\bar{t}}$. Moreover, when $\Theta=\pi/2$,
$E_L$ decreases rapidly with $M_{t\bar{t}}$. On the other hand, when $M_{t\bar{t}}$ is fixed to be $1000$ GeV, $E_L$ slowly increases but experiences a significant drop with $\Theta$. The smallest value of $E_L$ is attained at $(\Theta=\pi/2, M_{t\bar{t}}=1000$ GeV).

In Fig. 1(b), $E_R=\log_2\frac{1}{c}+H(A|B)=1+H(A|B)$ is illustrated for $c=1/2$. $E_R$ fluctuates with $\Theta$ and $M_{t\bar{t}}$, where for most of the cases, $E_R$ is $1$. However, in a small parameter region, $E_R$ decreases as $\Theta$ or $M_{t\bar{t}}$ increases. The smallest value of $E_R$ occurs at the same point as $E_L$, namely, $\Theta=\pi/2$ and $M_{t\bar{t}}=1000$ GeV. When the value of $M_{tt}$ increases, the resulting values become smaller. At the limit that $M_{tt}$ approaches to infinity, both $E_L$ and $E_R$ are equal to zero.

The tightness $\Delta_E=E_L-E_R$ is presented in Fig. 1(c), which is always greater than or equal to zero. Specifically, $\Delta_E=0$ only when $\Theta=0$ and $\pi/2$.
This indicates that the bound is state dependent. It is saturated for the pairs of observables $\{\sigma_x, \sigma_z\}$. For fixed $M_{t\bar{t}}$, $\Delta_E$ first increases and reaches a maximum value, and then decreases
with increase of $\Theta$.

\begin{figure}[htbp!]
\centering
\subfigure[]{
\includegraphics[width=.30\textwidth]{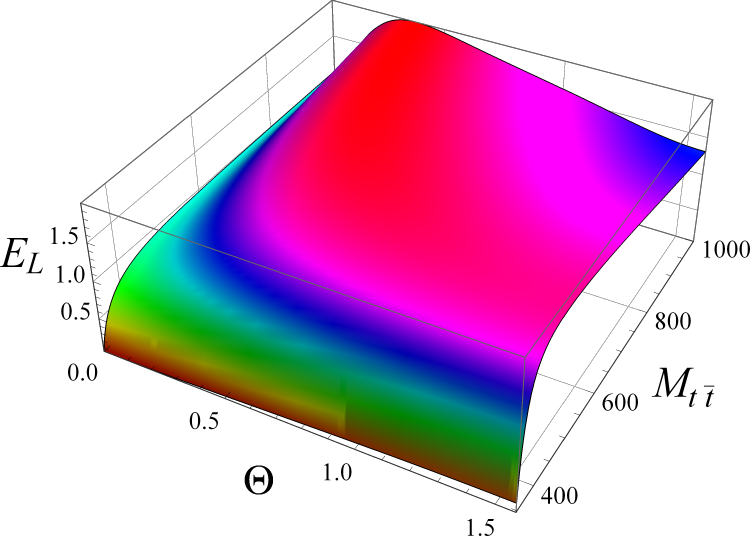}
}
\subfigure[]{
\includegraphics[width=.30\textwidth]{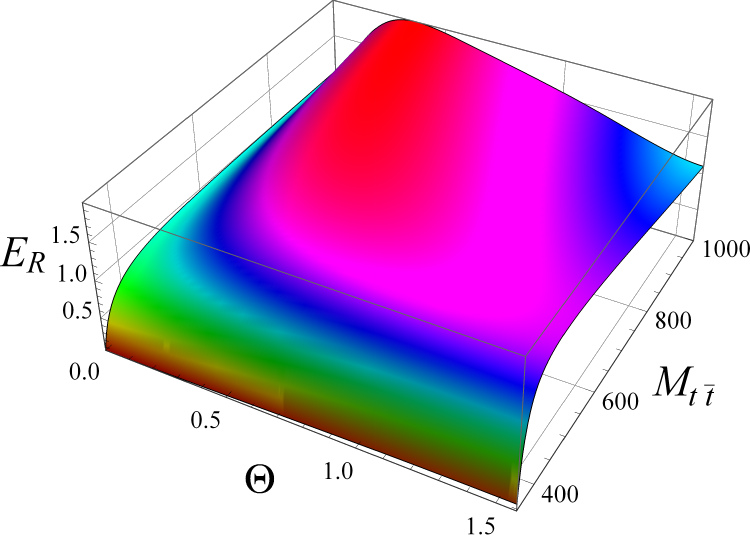}
}
\subfigure[]{
\includegraphics[width=.30\textwidth]{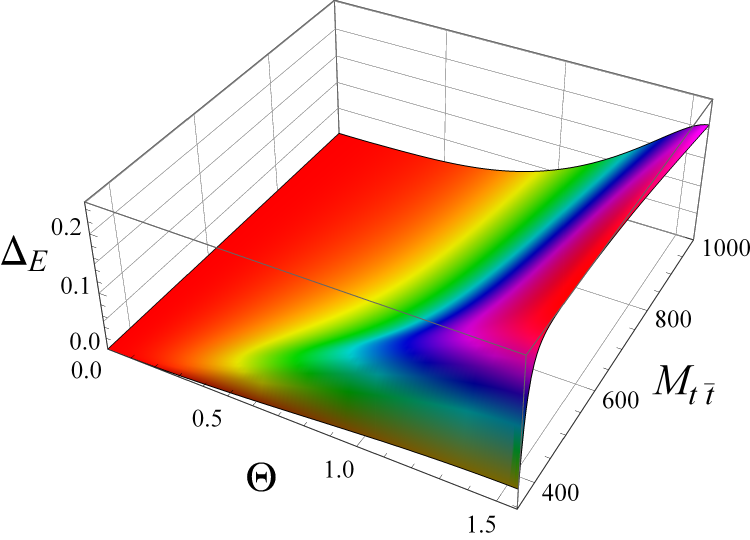}
}
\caption{Entropic uncertainty relation for the production of quarks via gluons ($gg\rightarrow t\bar{t}$). (a) shows the variation of $E_L$ with $\Theta$ and $M_{t\bar{t}}$. (b) shows the variation of $E_R$ with $\Theta$ and $M_{t\bar{t}}$, and (c) shows the variation of $\Delta_E$ with $\Theta$ and $M_{t\bar{t}}$. The unit of $M_{t\bar{t}}$ is GeV.
}\label{f2}
\end{figure}

Fig. \ref{f2} illustrates the entropic uncertainty relation for the production of quarks via gluons ($gg\rightarrow t\bar{t}$). The subfigures (a), (b) and (c) display the variations of $E_L$, $E_R$ and $\Delta_E$ as functions of $\Theta$ and $M_{t\bar{t}}$, respectively. In Fig. 2(a), $E_L$ exhibits a progressive increase with $\Theta$ when $M_{t\bar{t}}=346$ GeV. However, for $M_{t\bar{t}}=1000$ GeV, $E_L$ initially increases and then decreases smoothly with $\Theta$. For any fixed $\Theta$, $E_L$ grows with the increase of $M_{t\bar{t}}$.

Fig. \ref{f2}(b) shows that $E_R$ behaves similarly to $E_L$. For any fixed $\Theta$, $E_R$ also increases with $M_{t\bar{t}}$. Fig. \ref{f2}(c) presents the tightness $\Delta_E$. At $\Theta=0$, $\Delta_E$ is zero, indicating that maximum tightness. When $\Theta=\pi/2$, $\Delta_E$ initially increases rapidly and then expands gradually with the increasing $M_{t\bar{t}}$. The bound is dependent on the state, and saturated when $\Theta=0$ for pairs of observables $\{\sigma_x, \sigma_z\}$. However, it is not saturated at $\Theta=\pi/2$. The point $(\Theta=\pi/2, M_{t\bar{t}}=1000$ GeV) represents the maximum loss of tightness. For $M_{t\bar{t}}=346$ GeV, $\Delta_E$ increases quickly with $\Theta$. When $M_{t\bar{t}}$ is set to be $1000$ GeV, $\Delta_E$ gradually increases with $\Theta$.

\begin{figure}[htbp!]
\centering
\subfigure[]{
\includegraphics[width=.40\textwidth]{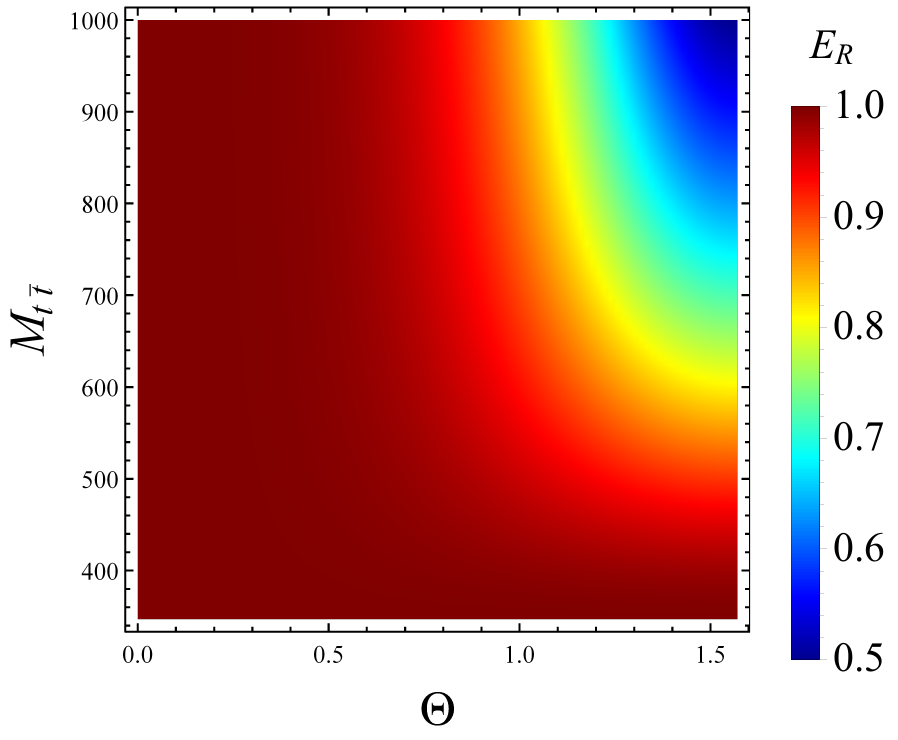}
}
\subfigure[]{
\includegraphics[width=.40\textwidth]{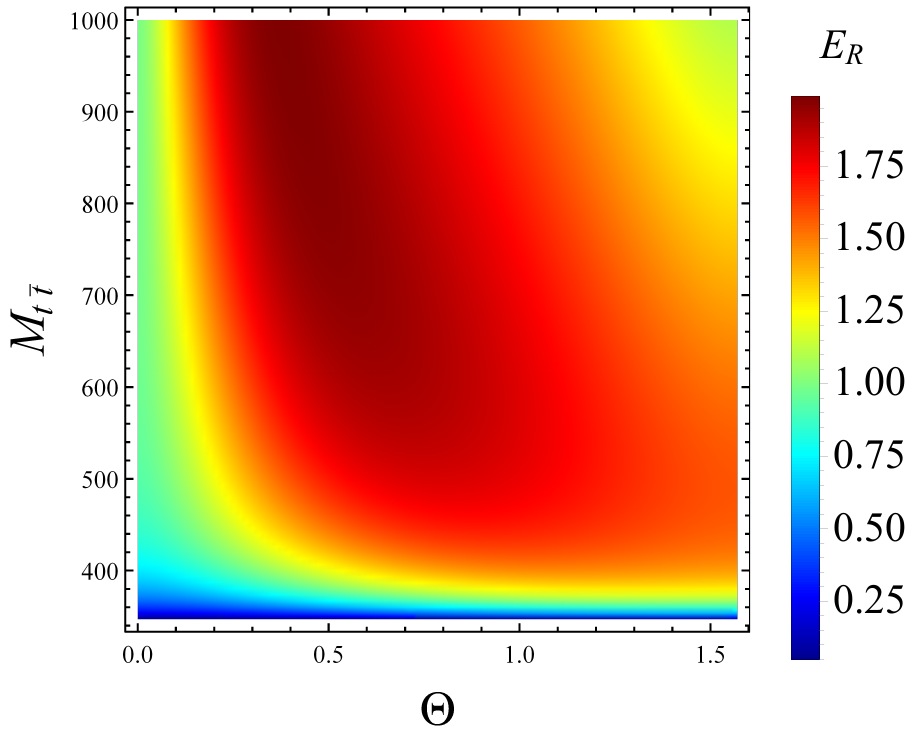}
}
\caption{Entropic uncertainty relation for the production of top quarks via quark and gluons initiated channels. (a) shows the variation of $E_R$ with $\Theta$ and $M_{t\bar{t}}$ via quarks ($q\bar{q}\rightarrow t\bar{t}$). (b) shows the variation of $E_R$ with $\Theta$ and $M_{t\bar{t}}$ via gluons ($gg\rightarrow t\bar{t}$). The unit of $M_{t\bar{t}}$ is GeV.
}\label{fer}
\end{figure}
The bound $E_R$ of entropic uncertainty relation witnesses entanglement when $E_R<1$. The production of top quarks through quark and gluon initiated channels is illustrated in Fig. \ref{fer}. Fig. \ref{fer}(a) displays the variation of $E_R$ with respect to $\Theta$ and $M_{t\bar{t}}$ via quarks ($q\bar{q}\rightarrow t\bar{t}$). The maximum entanglement is achieved at $(\Theta=\pi/2, M_{t\bar{t}}=1000$ GeV). Meanwhile, Fig. \ref{fer}(b) indicates the variation of $E_R$ with $\Theta$ and $M_{t\bar{t}}$ via gluons ($gg\rightarrow t\bar{t}$). However, the maximum entanglement occurs at $(\Theta=0, M_{t\bar{t}}=346$ GeV).

\begin{figure}[htbp!]
\centering
\subfigure[]{
\includegraphics[width=.40\textwidth]{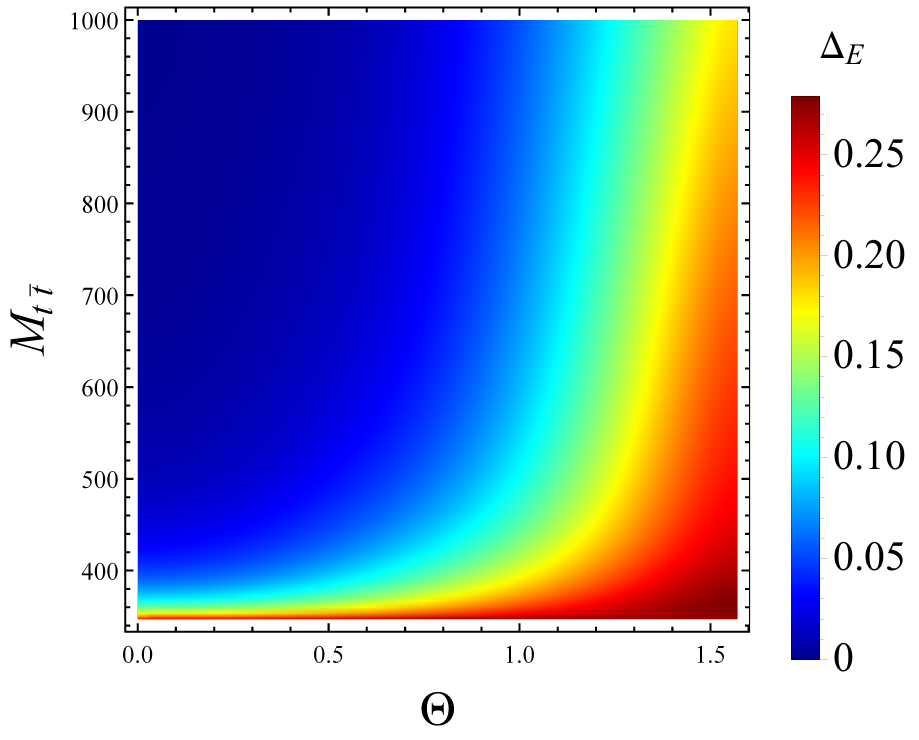}
}
\subfigure[]{
\includegraphics[width=.40\textwidth]{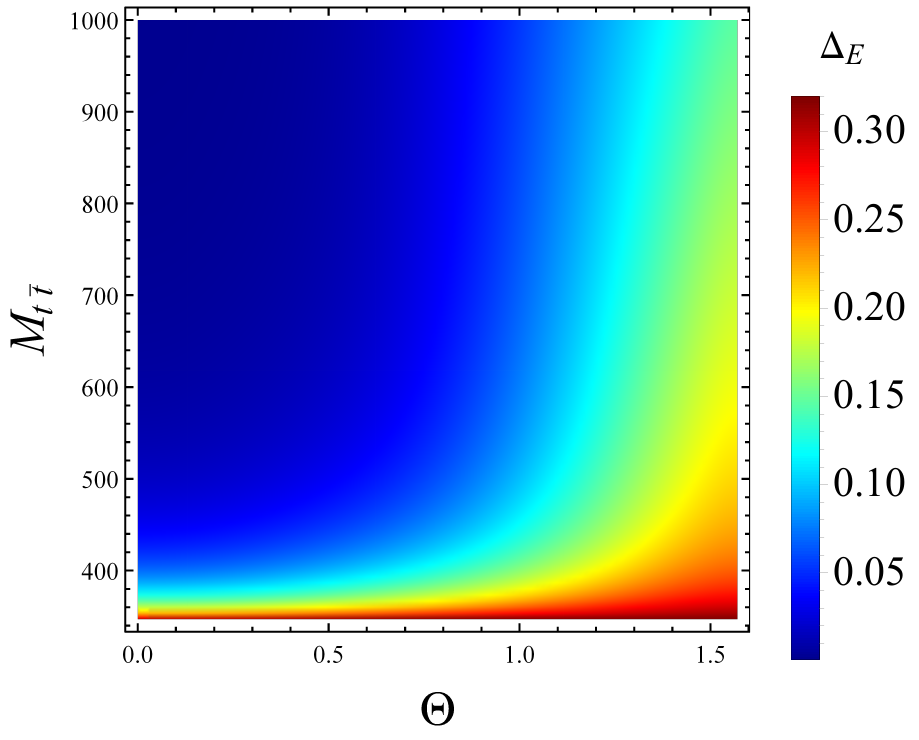}
}
\subfigure[]{
\includegraphics[width=.40\textwidth]{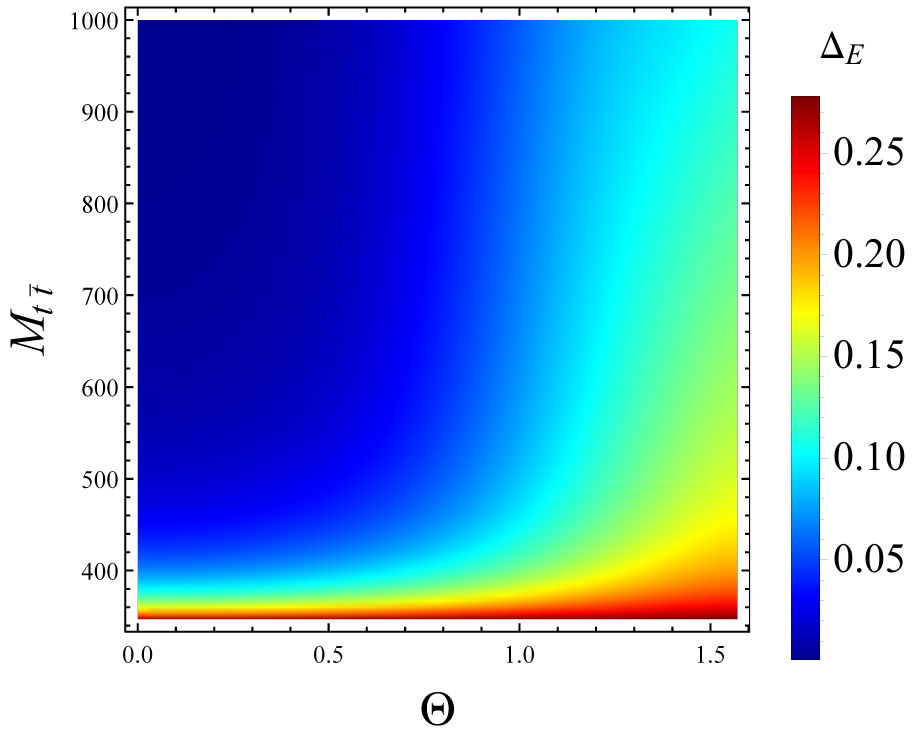}
}
\subfigure[]{
\includegraphics[width=.40\textwidth]{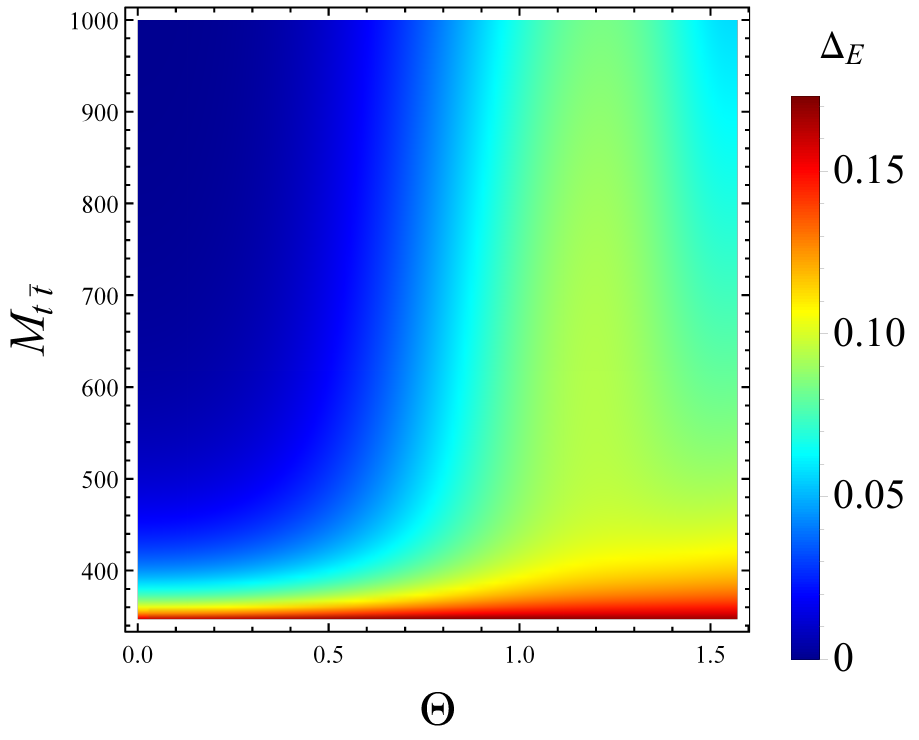}
}
\caption{Entropic uncertainty relation for quark and gluon initial state mixing
for varied probability $w_I$. The density of $\Delta_E$ vs $\Theta$ and $M_{t\bar{t}}$: (a) $w_{q\bar{q}}=0.2$; (b) $w_{q\bar{q}}=0.4$; (c) $w_{q\bar{q}}=0.6$; (d) $w_{q\bar{q}}=0.8$.  The unit of $M_{t\bar{t}}$ is GeV.
}
\label{f3}
\end{figure}
Fig. \ref{f3} displays the density of $\Delta_E$ for the mixing of quarks and gluons initiated channels as a function of $\Theta$ and $M_{t\bar{t}}$, with varied probabilities $w_{q\bar{q}}=0.2, 0.4, 0.6, 0.8$. The subfigures (a), (b), (c) and (d) correspond to the mixing probabilities 0.2, 0.4, 0.6, and 0.8, respectively. They exhibit a similar trend in terms of tightness across the varying mixing probabilities. As $\Theta$ increases, the tightness also increases. However, as $M_{t\bar{t}}$ increases, $\Delta_E$ exhibits a different behavior, i.e., it decreases instead. In certain parameter regions, the tightness vanishes. The maximum value of tightness is observed at $\Theta=\pi/2$ and $M_{t\bar{t}}=346$ GeV. In Fig. \ref{f3}(d), different from the subfigures (a)-(c), a slight yellow peak appears. This indicates that when $M_{t\bar{t}}=1000$ GeV, $\Delta_E$ initially increases and then decreases with respect to $\Theta$. While when $M_{t\bar{t}}=346$ GeV, the behavior is similar to that of subfigures (a)-(c), with a slow increase as $\Theta$ increases.

\begin{figure}[htbp!]
\centering
\subfigure[]{
\includegraphics[width=.40\textwidth]{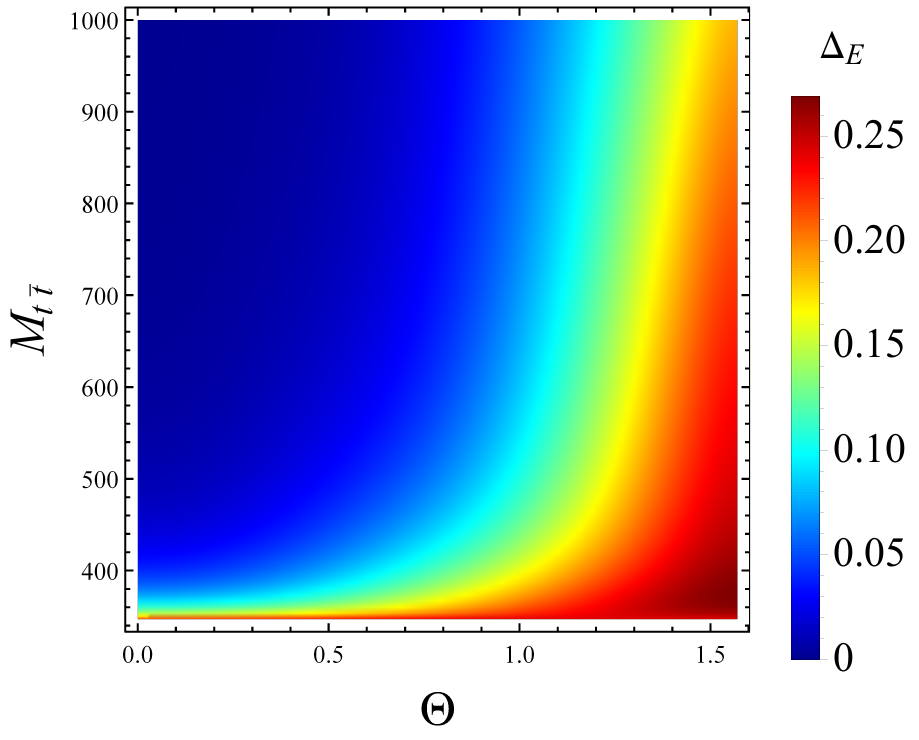}
}
\subfigure[]{
\includegraphics[width=.40\textwidth]{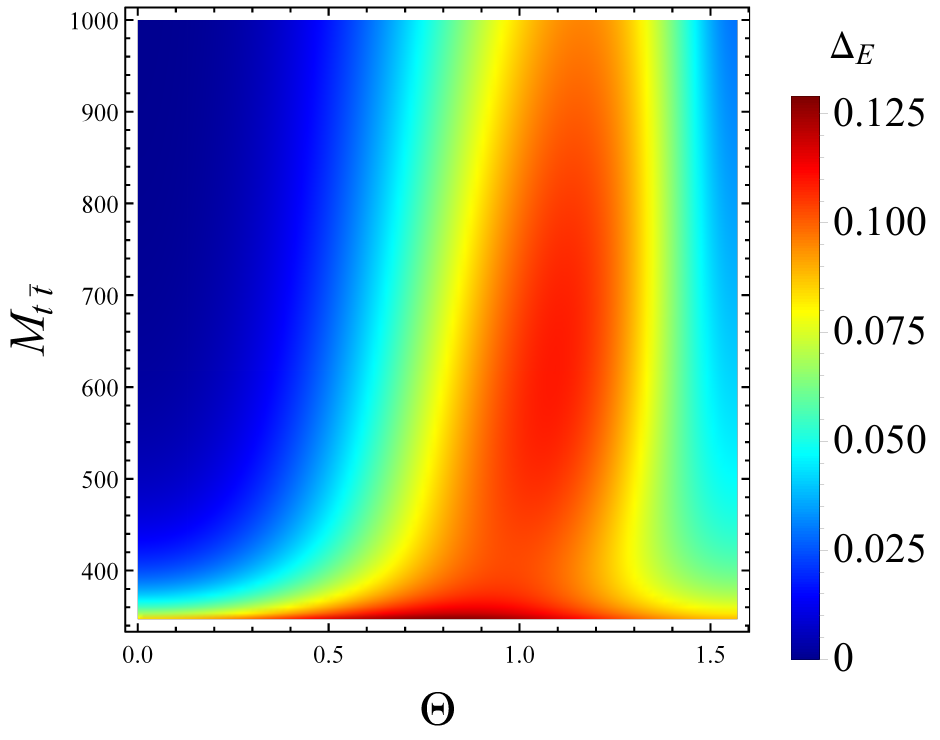}
}
\caption{Entropic uncertainty relation for $t\bar{t}$ production at the LHC for $\sqrt{s}=13$ TeV (a). $t\bar{t}$ production at the Tevatron for $\sqrt{s}=2$ TeV (b). The unit of $M_{t\bar{t}}$ is GeV.
}
\label{flt1}
\end{figure}
The entropic uncertainty relation of two specific hadronic processes can be observed in Fig. \ref{flt1}. One process refers to $pp$ collisions at a center-of-mass energy of $\sqrt{s}=13$ TeV, which corresponds to Run 2 at the LHC (a). Another process involves $p\bar{p}$ collisions at a center-of-mass energy of $\sqrt{s}=2$ TeV, which closely approximates the actual value of $\sqrt{s}=1.96$ TeV at the Tevatron (b) \cite{Afik2022}.

\begin{figure}[htbp!]
\centering
\subfigure[]{
\includegraphics[width=.40\textwidth]{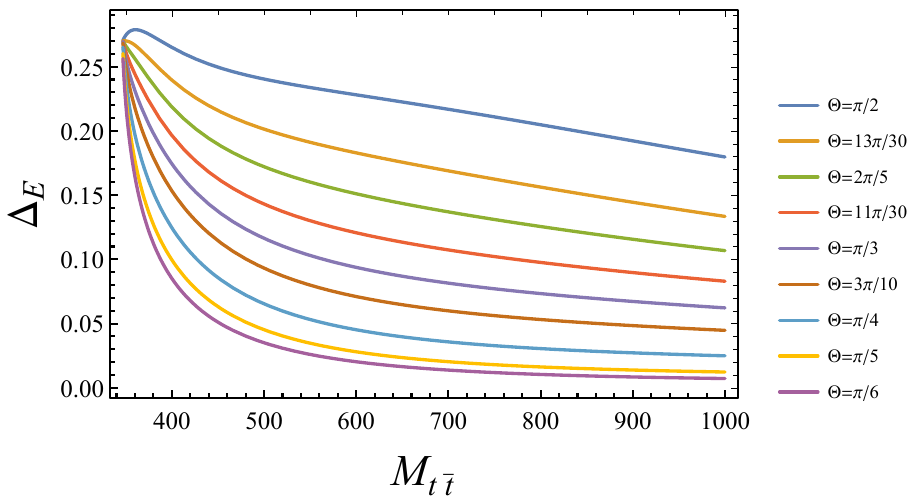}
}
\subfigure[]{
\includegraphics[width=.40\textwidth]{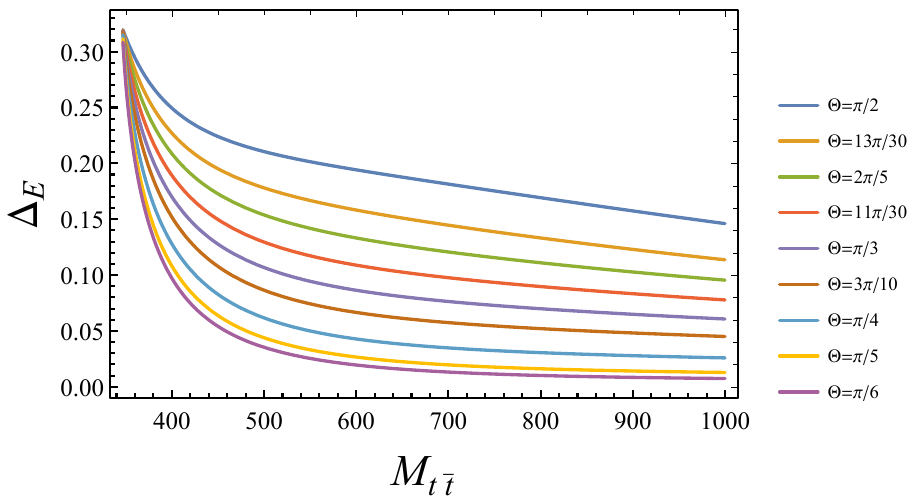}
}
\subfigure[]{
\includegraphics[width=.40\textwidth]{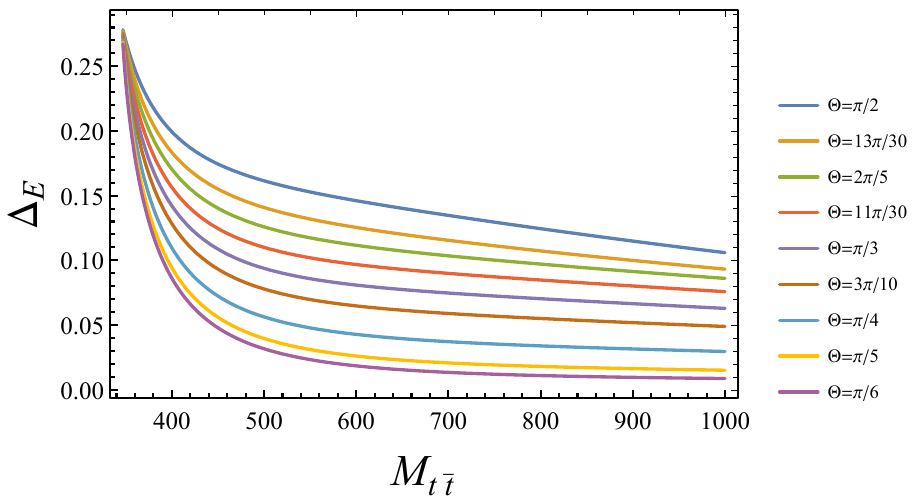}
}
\subfigure[]{
\includegraphics[width=.40\textwidth]{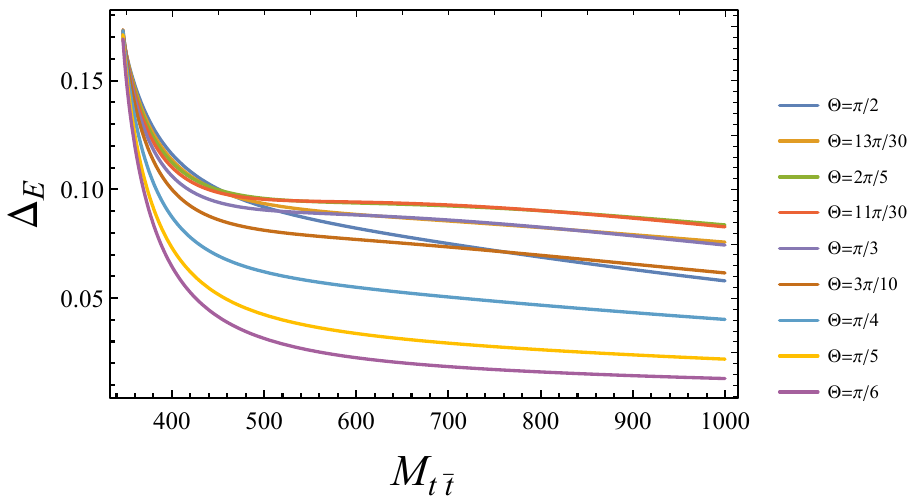}
}
\caption{Entropyic uncertainty relation for quark and gluon initial state mixing. The density of $\Delta_E$ vs $M_{t\bar{t}}$ for $\Theta=\pi/2, 13\pi/30, 2\pi/5, 11\pi/30, \pi/3, 3\pi/10, \pi/4, \pi/5, \pi/6$, with different probabilities: (a) $w_{q\bar{q}}=0.2$, (b) $w_{q\bar{q}}=0.4$, (c) $w_{q\bar{q}}=0.6$, (d) $w_{q\bar{q}}=0.8$. The unit of $M_{t\bar{t}}$ is $GeV$.
}
\label{f4}
\end{figure}
Fig. \ref{f4} presents the slice figure for Fig. \ref{f3} by displaying the 2D plots of $\Delta_E$ with varied mixing probabilities $w_{q\bar{q}}=0.2,0.4,0.6,0.8$, corresponding to figures (a), (b), (c), (d) for the respective $\Theta$ values $\pi/2, 13\pi/30, 2\pi/5, 11\pi/30, \pi/3,3\pi/10, \pi/4, \pi/5, \pi/6$, with different colored lines correspond to different $\Theta$ values. From Fig. \ref{f4}(a), (b), (c) and (d), it can be observed that as $\Theta$ decreases, the majority of $\Delta_E$ decreases with respect to $M_{t\bar{t}}$. However, for the mixing probability $w_{q\bar{q}}=0.8$, there is not a strictly monotonic relationship between $\Delta_E$ and $M_{t\bar{t}}$ as $\Theta$ increases. In Fig. \ref{f4}(a), there is also a small parameter region of $M_{t\bar{t}}$ where $\Delta_E$ initially increases and then declines slowly with $M_{t\bar{t}}$ for $\Theta=\pi/2$ and $\Theta=13\pi/30$. In Fig. \ref{f4}(d), one sees that $\Delta_E$ decreases more rapidly with $M_{t\bar{t}}$ compared to the other cases for $\Theta=\pi/2$.

Next we investigate the relationship between quantum Fisher information and the generating angle $\Theta$ and mass $M_{t\bar{t}}$ associated with the generation of $t\bar{t}$ pairs by quarks and gluons.

\begin{figure}[htbp!]
\centering
\subfigure[]{
\includegraphics[width=.40\textwidth]{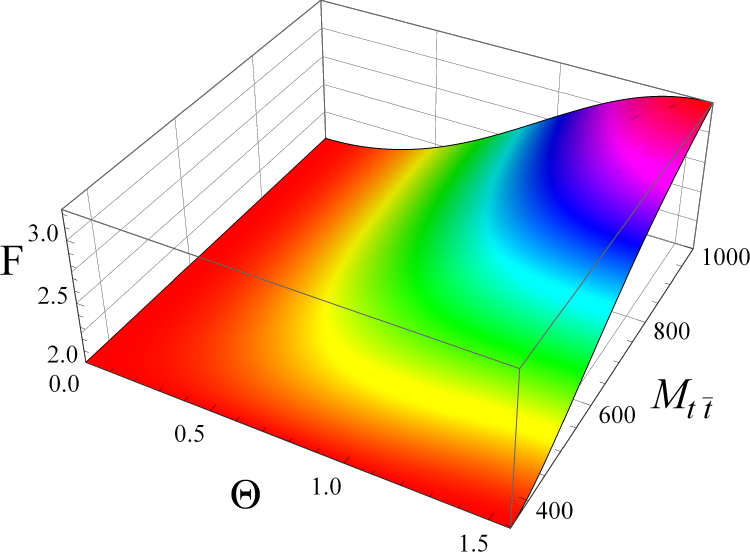}
}
\subfigure[]{
\includegraphics[width=.40\textwidth]{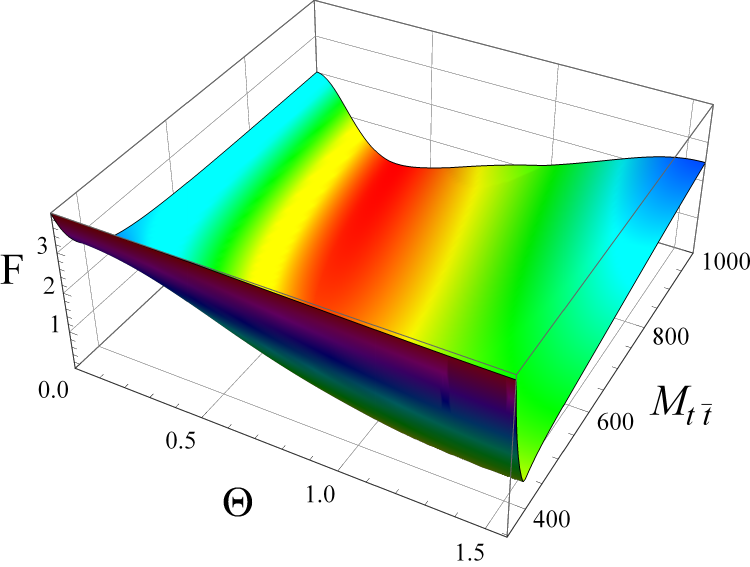}
}
\caption{Quantum Fisher information versus $\Theta$ and $M_{t\bar{t}}$ for (a) quarks, and (b) gluons.
}
\label{f5}
\end{figure}

Fig. \ref{f5} shows the quark and gluon contributions to the production of $t\bar{t}$ pairs. The variations of the quantum Fisher information ($F$) as functions of $\Theta$ and $M_{t\bar{t}}$ are shown in Fig. \ref{f5}(a) for quarks and in Fig. \ref{f5}(b) for gluons. One sees in Fig. \ref{f5}(a) that $F$ is larger than 2 for large cases. However, when $\Theta=\pi/2$, $F$ increases slowly with the increase of $M_{t\bar{t}}$. Specifically, when $M_{t\bar{t}}=1000$ GeV, $F$ also increases slowly with increase of $\Theta$. The maximum values of $F$ is attained at $(\Theta=\pi/2, M_{t\bar{t}}=1000$ GeV).

The behavior of $F$ for gluons in Fig. \ref{f5}(b) is more complex compared with quarks. When $M_{t\bar{t}}=346$, $F$ is near maximal value for almost $\Theta$. However, when $M_{t\bar{t}}=1000$ GeV, $F$ first decreases and reaches the valley floor, and then increases. The maximum value is attained at $(\Theta=0, M_{t\bar{t}}=346$ GeV). Moreover, $F$ falls and then grows with the increase of $M_{t\bar{t}}$ for given $\Theta=\pi/2$.

\begin{figure}[htbp!]
\centering
\subfigure[]{
\includegraphics[width=.40\textwidth]{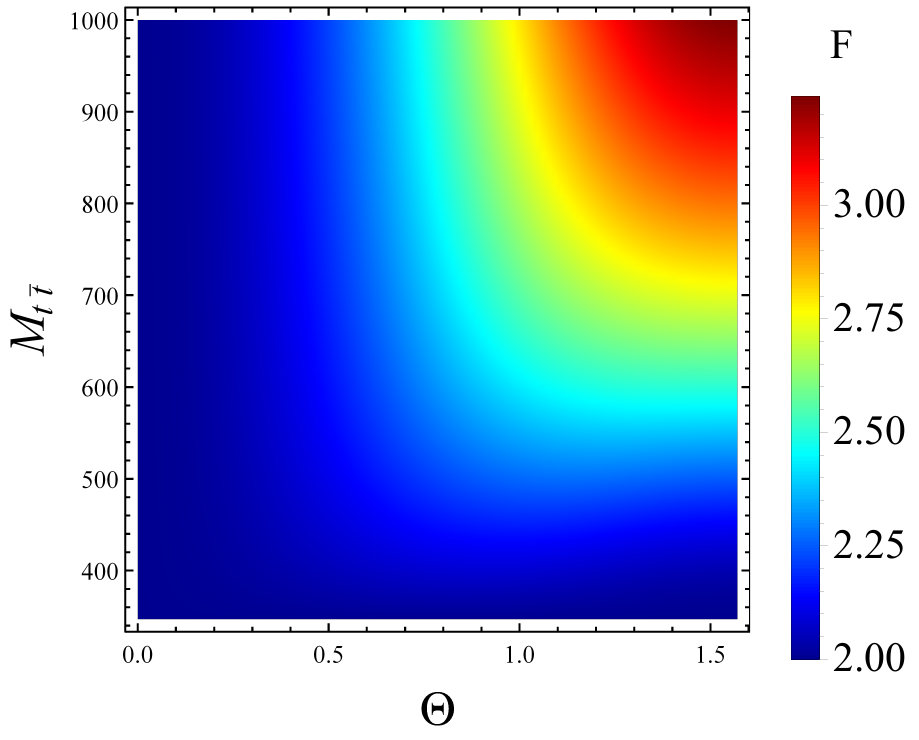}
}
\subfigure[]{
\includegraphics[width=.40\textwidth]{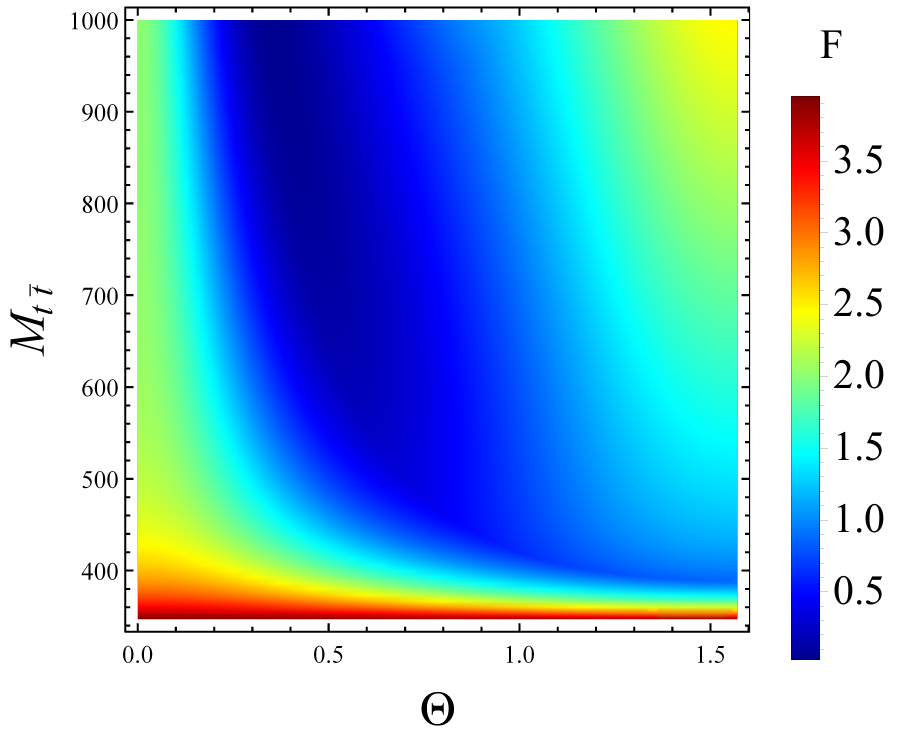}
}
\caption{Quantum Fisher information for quarks and gluons represented as density graphs. The density of $F$ is plotted against $\Theta$ and $M_{t\bar{t}}$ in (a) for quarks and in (b) for gluons, respectively.}
\label{f6}
\end{figure}

Fig. \ref{f6}(a) and Fig. \ref{f6}(b) illustrate the density plots for quarks and gluons, respectively, in the generation of $t\bar{t}$ pairs as in Figure \ref{f5}. Both the quark and gluon density plots show the same color contour for the $F$, resulting in the generation of $t\bar{t}$. The quarks' prominent color region with $F>2$ exhibits entanglement characteristics, similar to those shown in Figure 3(b) of reference \cite{Afik2021}. The range of values $F$ for gluons also resembles the concurrence depicted in Figure 3(a) of reference \cite{Afik2021}. The quarks attain their maximum values at $(\Theta=\pi/2, M_{t\bar{t}}= 1000$ GeV). In contrast, the gluons reach their maximum values at $(\Theta=0, M_{t\bar{t}}= 346$ GeV).

\begin{figure}[htbp!]
\centering
\subfigure[]{
\includegraphics[width=.40\textwidth]{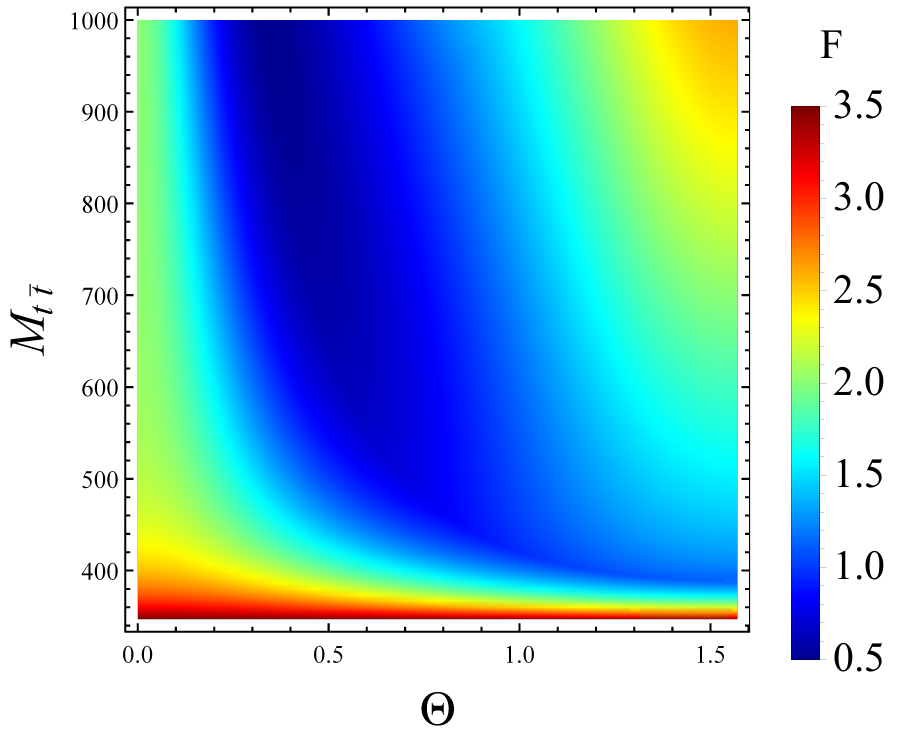}
}
\subfigure[]{
\includegraphics[width=.40\textwidth]{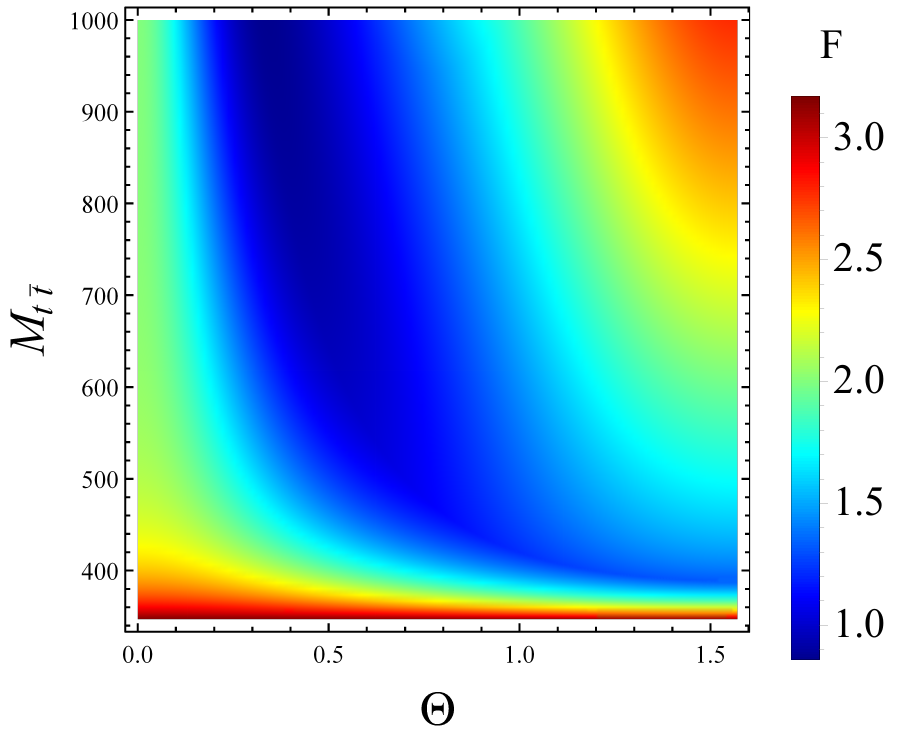}
}
\subfigure[]{
\includegraphics[width=.40\textwidth]{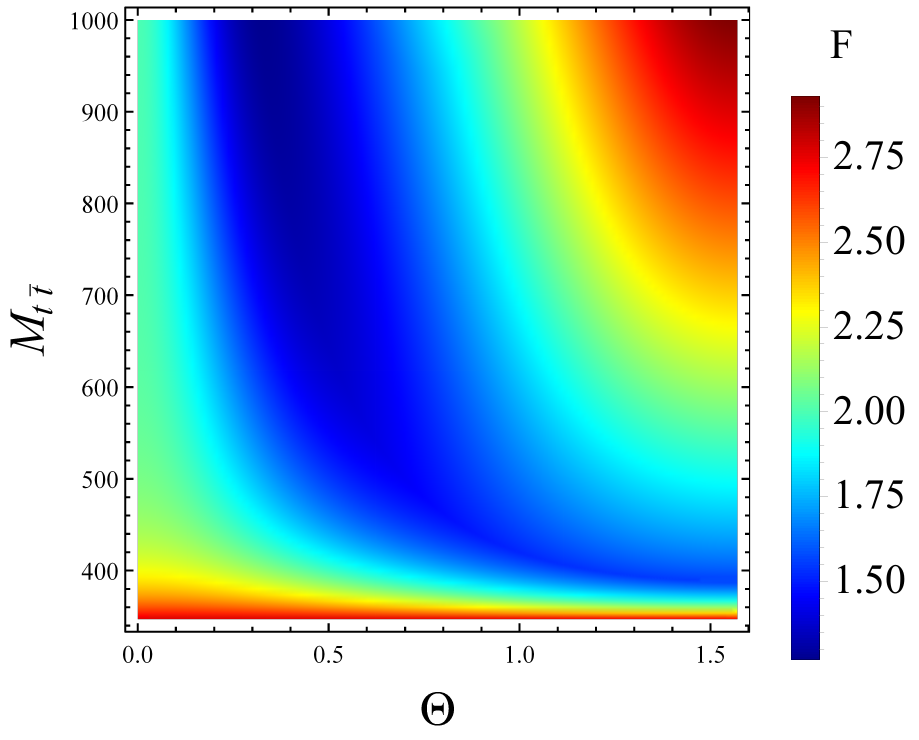}
}
\subfigure[]{
\includegraphics[width=.40\textwidth]{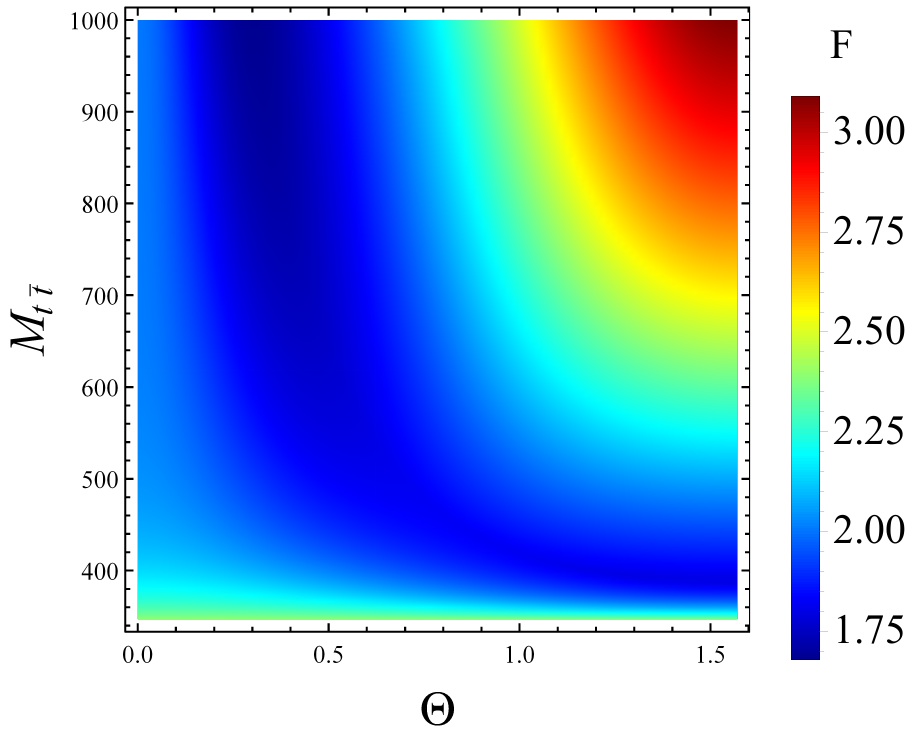}
}
\caption{Quantum Fisher information $F$ with respect to the mixing of quarks and gluons
initial state in different probabilities. (a) $w_{q\bar{q}}=0.2$, (b) $w_{q\bar{q}}=0.4$,
(c) $w_{q\bar{q}}=0.6$, (d) $w_{q\bar{q}}=0.8$.
}
\label{f7}
\end{figure}

We present the representation of the mixture of quarks and gluons process to generate the $t\bar{t}$ pair in Fig. \ref{f7}. The quantum Fisher information $F$ varies as a function of $\Theta$ and $M_{t\bar{t}}$. The subfigures (a), (b), (c) and (d) correspond to the probabilities $w_{q\bar{q}}=0.2, 0.4, 0.6,$ and $0.8$, respectively. As the degree $w_{q\bar{q}}$ of the mixing increases, the right corner red region expands but left corner red zone decreases, and the value of $F$ also changes. In the subfigures (a) and (b), the maximum value of $F$ is attained at $(\Theta=0, M_{t\bar{t}}=346$ GeV). However, in the subfigures (c) and (d), the maximum value of $F$ is attained at $(\Theta=\pi/2, M_{t\bar{t}}=1000$ GeV).

\begin{figure}[htbp!]
\centering
\subfigure[]{
\includegraphics[width=.40\textwidth]{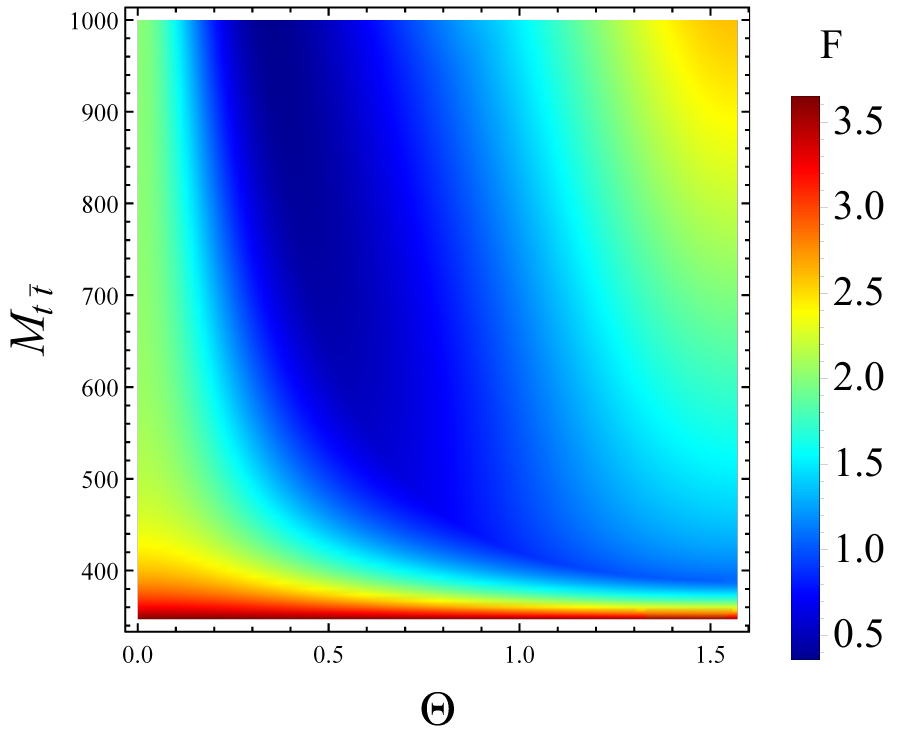}
}
\subfigure[]{
\includegraphics[width=.40\textwidth]{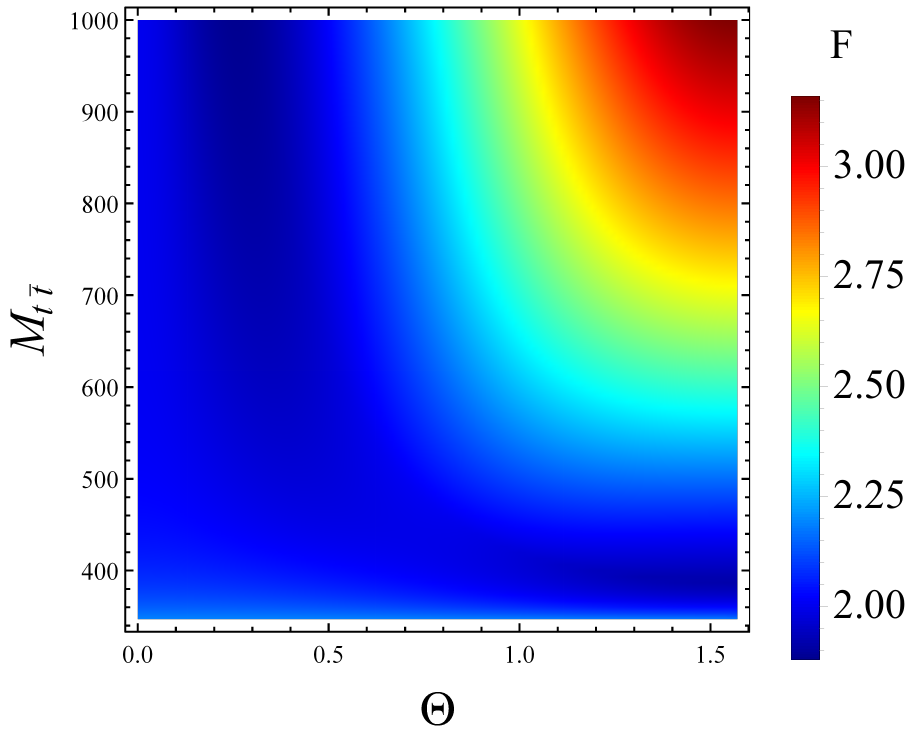}
}
\caption{Quantum Fisher information for $t\bar{t}$ production at the LHC for $\sqrt{s}=13$ TeV (a). $t\bar{t}$ production at the Tevatron for $\sqrt{s}=2$ TeV (b). The unit of $M_{t\bar{t}}$ is GeV.
}
\label{flt2}
\end{figure}
The quantum Fisher information of two specific hadronic processes is illustrated in Fig. \ref{flt2}. One process refers to $pp$ collisions at a center-of-mass energy of $\sqrt{s}=13$ TeV, corresponding to Run 2 at the LHC (a). Another process pertains to $p\bar{p}$ collisions at $\sqrt{s}=2$ TeV, which is in close proximity to the actual value of $\sqrt{s}=1.96$ TeV at the Tevatron accelerator (b) \cite{Afik2022}.

\begin{figure}[htbp!]
 \centering
 \subfigure[]{
\includegraphics[width=.40\textwidth]{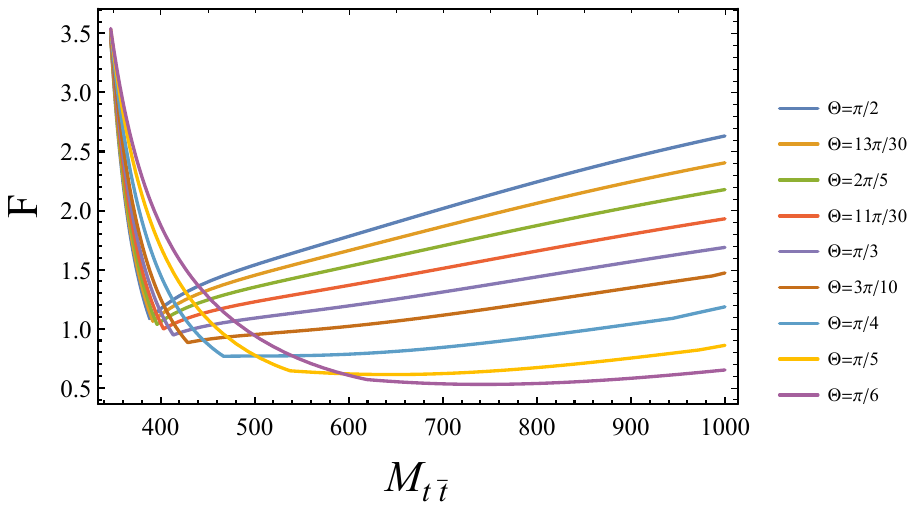}
}
\subfigure[]{
\includegraphics[width=.40\textwidth]{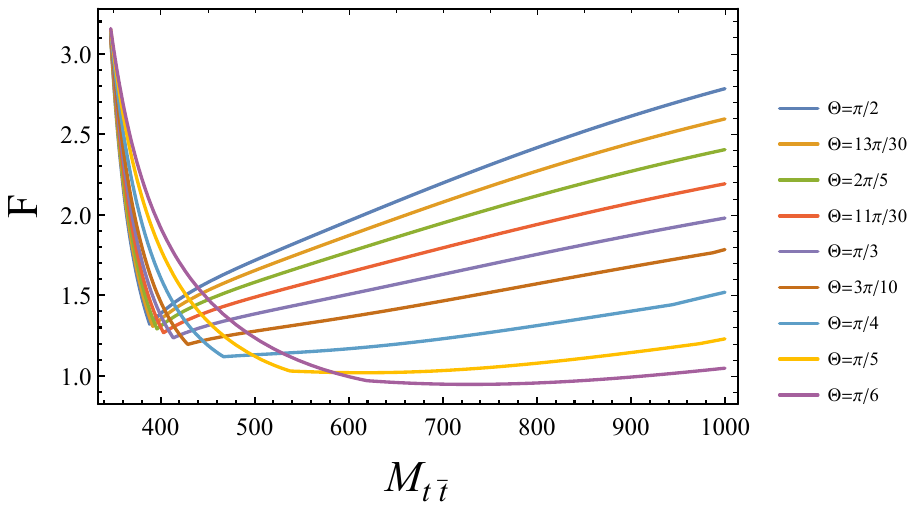}
}
\subfigure[]{
\includegraphics[width=.40\textwidth]{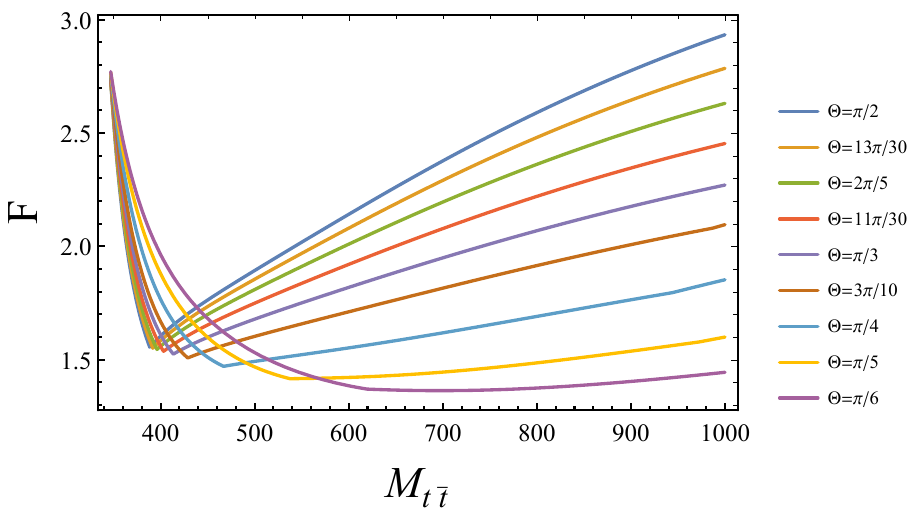}
}
\subfigure[]{
\includegraphics[width=.40\textwidth]{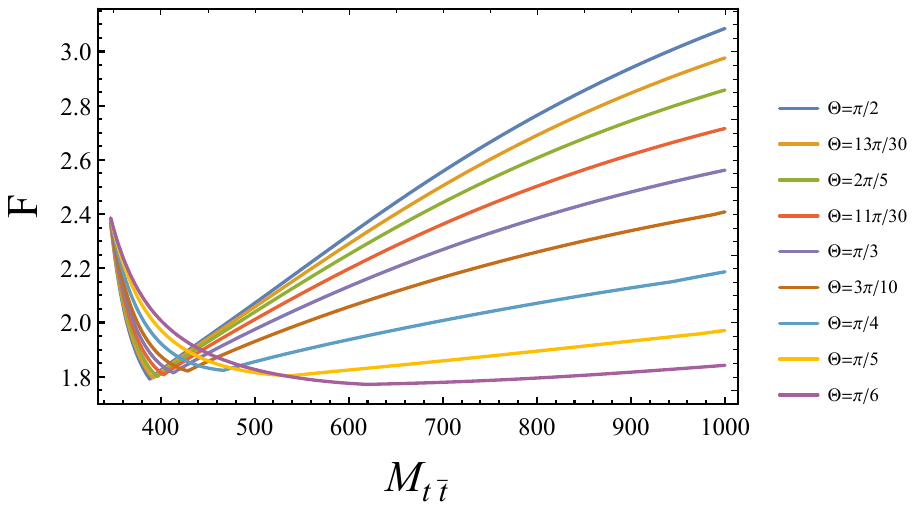}
}
\caption{Quantum Fisher information vs $M_{t\bar{t}}$ and mixing degree for the mixing of quarks and gluons initial state with respect to different probabilities: (a) $w_{q\bar{q}}=0.2$, (b) $w_{q\bar{q}}=0.4$, (c) $w_{q\bar{q}}=0.6$ and (d) $w_{q\bar{q}}=0.8$, for $\Theta=\pi/2, 13\pi/30, 2\pi/5, 11\pi/30, \pi/3, 3\pi/10, \pi/4, \pi/5, \pi/6$, respectively.
}
\label{f8}
\end{figure}
We plot the quantum Fisher information $F$ against the top quark-antiquark invariant mass $M_{t\bar{t}}$ for various mixing probabilities $w_{q\bar{q}}$ and fixed values of $\Theta=\pi/2, 13\pi/30, 2\pi/5, 11\pi/30, \pi/3, 3\pi/10,\\
\pi/4, \pi/5, \pi/6$ in Fig. \ref{f8}, which correspond to specific slices of Fig. \ref{f7}. One observes that $F$ falls and then grows with $M_{t\bar{t}}$ for all the four subfigures. However, for $\Theta=\pi/6$, $F$ decreases first and then increases slowly with $M_{t\bar{t}}$. Interestingly, for (a) and (b), the maximum value of $F$ happens at $(\Theta=0, M_{t\bar{t}}=346$ GeV). However, the maximum value of $F$ occurs at $(\Theta=\pi/2, M_{t\bar{t}}=1000$ GeV) for (c) and (d).

\section{Conclusions}
We have investigated the production of top quark pairs ($t\bar{t}$) through interactions involving quarks and gluons processes. Our analysis focuses on the behaviors exhibited by the system at the large hadron collider from the perspective of entropic uncertainty relations and quantum Fisher information. Notably, we have found that the entropic uncertainty relations are most stringent, while the quantum Fisher information is maximized when considering the production angle and mass of the particles. Our study may contribute significantly to a deeper comprehension of the LHC's workings and to the investigation of new physics in the realm of quarks.

\medskip

\noindent {\bf Acknowledgments}
This work is supported by the National Natural Science Foundation of China (NSFC) under Grant Nos. 12075159, 12171044 and 11905131, the specific research fund of the Innovation Platform for Academicians of Hainan Province. This work was also supported by Jiangxi Provincial Natural Science Foundation
under Grant No. 20224BAB201027.


%

\end{document}